\documentclass{article}

\usepackage{microtype}
\usepackage{graphicx}
\usepackage{booktabs} 
\usepackage{hyperref}
\usepackage{amsmath}
\usepackage{amssymb}
\usepackage{graphics}
\usepackage{graphicx}
\usepackage{subfig}
\usepackage{listings}
\usepackage{fullpage}
\usepackage{authblk}

\def\d{d} 

\def\C{{C}}

\def\bp{\noindent {\it Proof.}\ }
\def\ep{\hfill $\Box$} 

\def\be{\begin{equation}}
\def\ee{\end{equation}}

\newtheorem{prop}{Proposition}

\begin{document}

\title{Hierarchical Graph Clustering using Node Pair Sampling}
\author[1]{Thomas Bonald}
\author[1]{Bertrand Charpentier\thanks{Part of this work has been done while Bertrand Charpentier was a Master student at KTH, Sweden.}}
\author[2]{Alexis Galland}
\author[2]{Alexandre Hollocou}
\affil[1]{Telecom ParisTech, Paris, France}
\affil[2]{Inria, Paris, France}
\date{\today}

\maketitle

\begin{abstract}
We present a novel hierarchical graph clustering algorithm  inspired by modularity-based clustering techniques. The algorithm is agglomerative and based on a simple distance between clusters induced by the probability of sampling node pairs. We prove that this distance is reducible, which enables the use of the nearest-neighbor chain to speed up the agglomeration. The output of the algorithm is a regular  dendrogram, which reveals the multi-scale structure of the graph. The results are illustrated  on both synthetic and real datasets.
\end{abstract}

\section{Introduction}

Many datasets can be  represented as graphs, being the graph explicitely embedded in data (e.g., the friendship relation of a social network) or built through some suitable similarity measure between data items (e.g., the number of papers co-authored  by   two researchers). Such graphs often exhibit a complex, multi-scale community structure where each node is invoved in many groups of nodes, so-called communities, of different sizes. 

One of the most popular graph clustering algorithm is known as  
Louvain 
in name of the university of its inventors \cite{louvain}. It
is based on the greedy maximization 
of the {modularity}, a classical objective function introduced in \cite{newman2004}. The Louvain algorithm is fast, memory-efficient, and provides meaningful clusters in practice. It does not enable an analysis of the graph at different scales, however  \cite{fortunato2007,lancichinetti2011}. The resolution parameter available in the current version of the algorithm\footnote{See the  {\tt python-louvain} Python package.}  is not  directly related to  the number of clusters or the cluster sizes and thus is hard to adjust in practice. 

In this paper, we present a novel algorithm for  hierarchical  clustering that captures the multi-scale nature of real graphs. The algorithm is fast, memory-efficient and 
parameter-free.
It relies on 
a novel notion of  distance between clusters induced by the probability of sampling node pairs. We prove that this distance is reducible, which guarantees that the resulting hierarchical clustering can be represented by regular dendrograms and enables a fast implementation of our algorithm through the nearest-neighbor chain scheme, a classical  technique for agglomerative algorithms \cite{murtagh}. 

The rest of the paper is organized as follows. We  present the related work in Section \ref{sec:related}. The notation used in the paper, the distance between clusters used to aggregate nodes and the clustering algorithm are presented in Sections \ref{sec:not}, \ref{sec:dist} and \ref{sec:algo}. The link with  modularity and the Louvain algorithm is explained in Section \ref{sec:mod}. Section \ref{sec:exp} shows the experimental results and Section \ref{sec:conc} concludes the paper.

\section{Related work}
\label{sec:related}

Most graph clustering algorithms are not hierarchical and rely on some resolution parameter that allows  one to  adapt the clustering to the  dataset and to the intended purpose \cite{reichardt2006,arenas2008,lambiotte2015,newman2016}.
This parameter is hard to adjust in practice, which motivates the present work.

Usual
 hierarchical clustering techniques apply to vector data \cite{ward,murtagh}. They do not directly apply to graphs, unless the graph is embedded in some metric space, through  spectral techniques for instance  \cite{luxburg,donetti2004}. 
 
A number of 
 hierarchical clustering algorithms have been developped specifically for graphs. The most popular algorithms are {agglomerative} and  characterized by some {distance} between clusters, see \cite{newman2004fast,pons2005,huang2010,chang}. None of these distances has been proved to be reducible, a key property of our algorithm.
Among non-agglomerative algorithms,    the  divisive approach of \cite{newman2004} is based on the notion of edge betweenness while  the iterative approach of  \cite{sales2007} 
and \cite{lancichinetti2009}
look for local maxima of modularity  or some fitness function; other approaches rely on 
   statistical interence  \cite{clauset2008}, replica correlations \cite{ronhovde2009}
and graph wavelets \cite{tremblay2014}. To our knowledge, non of these algorithms has been proved to lead to regular dendrograms (that is, without inversion).

Finally, the Louvain algorithm also  provides a hierarchy,  induced by the successive aggregation steps  of the algorithm \cite{louvain}. This is not a full hierarchy, however, as there are typically a few aggregation steps.   Moreover,   the same resolution is used in the optimization of modularity across all levels of the hierarchy, while the numbers of clusters decrease rapidly after a few aggregation steps.  We shall see that our algorithm may be seen as a modified version of Louvain using a {\it sliding} resolution, that is adapted to the current agglomeration step of the algorithm. 

\section{Notation}
\label{sec:not}

Consider a weighted, undirected graph $G = (V,E)$ of $n$ nodes, with $V=\{1,\ldots,n\}$. Let $A$ be the corresponding weighted adjacency matrix. This is a symmetric, non-negative matrix such that for each $i,j \in V$, $A_{ij}> 0$ if and only if there is an edge between $i$ and $j$, in which case $A_{ij}$ is the weight of  edge $\{i,j\} \in E$. We refer to the weight of node $i$ as the sum of the weights of its incident edges, 
$$
w_i = \sum_{j\in V} A_{ij}.
$$
Observe that for unit weights, $w_i$ is the degree of node $i$.
The total weight of the nodes is:
$$
w =  \sum_{i\in V} w_{i} =  \sum_{i,j\in V}A_{ij}.
$$
We refer to a clustering $C$ as any partition of $V$. In particular,  each element of $C$ is a subset of $V$, we refer to as a {\it cluster}.

\section{Node pair sampling}
\label{sec:dist}

The  weights induce a probability distribution on node pairs,
$$
\forall i,j\in V,\quad  p(i,j) = \frac{A_{ij}}{w},
$$
and a probability distribution on nodes,
$$
\forall i\in V,\quad p(i) =\sum_{j\in V} p(i,j) =  \frac{w_i}{w}.
$$
Observe that the joint distribution $p(i,j)$ depends on the graph (in particular, only {\it neighbors} $i,j$ are sampled with positive probability), while the marginal distribution $p(i)$ depends on the graph through  the node weights only.

We define the {\it  distance} between two distinct nodes $i,j$ as the node pair sampling ratio\footnote{The distance $d$ is not a metric in general. We only require symmetry and non-negativity.}:
\be\label{eq:d}
\d(i,j) =  \frac{p(i)p(j)}{p(i,j)},
\ee
with $\d(i,j) = +\infty$ if $p(i,j) = 0$ (i.e., $i$ and $j$ are not neighbors). 
Nodes $i,j$ are {\it close} with respect to this distance if the pair $i,j$ is sampled much more frequently through the joint distribution $p(i,j)$ than through the product distribution $p(i)p(j)$. 
For unit weights, the joint distribution is uniform over the edges, so that 
the closest  node pair is the pair of  neighbors having the lowest degree product.

Another interpretation of the node distance $d$ follows from the conditional probability,
$$
\forall i,j\in V,\quad p(i|j) = \frac{p(i,j)}{p(j)} =  \frac{A_{ij}}{w_j}.
$$
This is the conditional probability of sampling $i$ given that  $j$ is sampled (from the joint distribution). The distance between $i$ and $j$ can then be written
$$
\d(i,j) =  \frac{p(i)}{p(i|j)} = \frac{p(j)}{p(j|i)}.
$$
Nodes $i,j$ are {\it close} with respect to this distance if  $i$  (respectively, $j$) is sampled much more frequently given that $j$ is sampled (respectively, $i$).

Similarly, consider a clustering $\C$ of the graph (that is, a partition of $V$).
The weights induce a probability distribution on cluster pairs,
$$
\forall a,b\in \C,\quad p(a,b) = \sum_{i\in a,j\in b} p(i,j),
$$ 
and a probability distribution on clusters,
$$
\forall a\in \C,\quad p(a) =  \sum_{i\in a} p(i) =\sum_{b\in \C} p(a,b).
$$
We define the {distance} between two distinct clusters $a,b$ as the cluster pair sampling ratio:
\be\label{eq:d2}
\d(a,b) =  \frac{p(a)p(b)}{p(a,b)},
\ee
with $\d(a,b) = +\infty$ if $p(a,b) = 0$ (i.e., there is no edge between clusters $a$ and $b$). 
Defining the conditional probability
$$
\forall a,b\in C,\quad p(a|b) = \frac{p(a,b)}{p(b)},
$$
which is the conditional probability of sampling $a$ given  that  $b$ is sampled,
we get
$$
\d(a,b) =  \frac{p(a)}{p(a|b)} = \frac{p(b)}{p(b|a)}.
$$
This distance will be used in the agglomerative algorithm to merge the closest clusters. We have the following key results.

\begin{prop}[Update formula]\label{prop:up}
For any distinct clusters $a,b,c\in\C$,
$$
d(a\cup b,c) = \left(\frac{p(a)}{p(a\cup b)}\frac 1 {d(a,c)} +\frac{p(b)}{p(a\cup b)}\frac 1 {d(b,c)} \right)^{-1}.
$$
\end{prop}
\bp
We have:
\begin{align*}
p(a\cup b) &p(c) d(a\cup b,c)^{-1} = p(a\cup b,c),\\
& = p(a,c) + p(b,c),\\
& = p(a) p(c) d(a,c)^{-1}
  + p(b) p(c) d(b,c)^{-1},
\end{align*}
from which the formula follows.
\ep

\begin{prop}[Reducibility]
For any distinct clusters $a,b,c\in\C$,
$$
d(a\cup b,c)\ge \min(d(a,c),d(b,c)).
$$
\end{prop}
\bp
By Proposition \ref{prop:up}, $d(a\cup b,c)$ is a weighted harmonic mean of $d(a,c)$ and $d(b,c)$, from which the inequality follows.
\ep

By the reducibility property, merging clusters $a$ and $b$ cannot decrease their minimum distance  to any other cluster $c$.

\section{Clustering algorithm}
\label{sec:algo}

The agglomerative approach consists in starting from individual clusters (i.e., each node is in its own cluster) and  merging clusters recursively. At each step of the algorithm, the two {\it closest} clusters are merged. We obtain the following algorithm:

\begin{enumerate}
\item {\bf Initialization} \\
$\C \gets \{\{1\} , \ldots , \{n\}\}$\\
$L\gets \emptyset$
\item {\bf Agglomeration}\\
 For $t = 1,\ldots,n - 1$,
\begin{itemize}
\item $a,b \gets \arg\min_{a',b'\in \C, a'\ne b'} d(a',b')$
\item $\C\gets \C \setminus \{a, b\}$
\item  $\C\gets \C\cup  \{a\cup b\}$
\item $L \gets L \cup \{\{a, b\}\}$
\end{itemize}
\item Return $L$
\end{enumerate}

 The successive clusterings $C_0,C_1,\ldots,C_{n-1}$ produced by the algorithm, with $C_0 =  \{\{1\} , \ldots , \{n\}\}$,  can be recovered from the list $L$ of successive merges. Observe that clustering $C_t$ consists of $n-t$ clusters, for $t=0,1,\ldots,n-1$. By the reducibility property,  the corresponding sequence of distances $d_0,d_1,\ldots,d_{n-1}$ between merged clusters, with $d_0 = 0$,  is non-decreasing, resulting in a regular dendrogram (that is, without inversions) \cite{murtagh}.

It is worth noting that the graph $G$ does not need to be connected. If the graph consists of $k$ connected components, then the clustering $C_{n-k}$ gives these $k$ connected components, whose respective distances are infinite; the $k-1$ last merges can then be done in an arbitrary order. Moreover, the hierarchies associated with these connected components are independent of one another (i.e., the algorithm successively applied to the corresponding subgraphs would produce exactly the same clustering). Similarly, we expect the clustering of weakly connected subgraphs to be approximately independent of one another.  This is not the case of the Louvain algorithm, whose clustering depends on the whole graph through the total  weight $w$,  a shortcoming related to the resolution limit of modularity  (see Section \ref{sec:mod}).

\paragraph{Aggregate graph.}
In view of \eqref{eq:d2}, for any clustering $\C$ of $V$, the distance $d(a,b)$ between two clusters $a,b\in C$ is the distance between two nodes $a,b$ of the following aggregate graph: nodes are the elements of $\C$ and the weight   between   $a,b\in \C$ (including the case $a=b$, corresponding to a self-loop) is $\sum_{i\in a,j\in b} A_{ij}$. Thus the agglomerative algorithm can be implemented by merging nodes and updating the weights (and thus the distances between nodes)   at each step of the algorithm. Since the initial nodes of the graph are indexed from $0$ to $n-1$, we index the  cluster created at step $t$ of the algorithm  by $n+t$. We obtain the following  version of the above algorithm, where the clusters are replaced by their respective indices:

\begin{enumerate}
\item {\bf Initialization}\\
 $V \gets \{1,\ldots,n\}$\\
  $L\gets \emptyset$
\item {\bf Agglomeration}\\
 For $t = 1,\ldots,n - 1$,
\begin{itemize}
\item $i,j \gets \arg\min_{i',j'\in V, i'\ne j'} d(i',j')$
\item $L \gets L \cup \{\{i, j\}\}$
\item $V\gets V \setminus \{i, j\}$
\item $V\gets V\cup  \{n + t\}$
\item $p(n+t) \gets p(i) + p(j)$
\item $p(n+t,u) \gets p(i,u) + p(j,u)$ for $u\in V\setminus\{n\cup t\}$
\end{itemize}
\item Return $L$
\end{enumerate}


\paragraph{Nearest-neighbor chain.} By the reducibility property of the distance, the algorithm can be implemented through the nearest-neighbor chain scheme \cite{murtagh}. Starting from an arbitrary node, a chain a nearest neighbors is formed. Whenever  two nodes of the chain are mutual nearest neighbors, these two nodes are  merged and the chain is updated recursively, until the initial node is eventually merged. This scheme reduces the search of a global minimum (the pair of nodes $i,j$ that minimizes $d(i,j)$) to that of a local minimum (any pair of nodes $i,j$ such that $d(i,j) = \min_{j'}d(i,j') = \min_{i'} d(i',j)$), which speeds up the algorithm while returning exactly the  {\it same} hierarchy.
It only requires a consistent tie-breaking rule for equal distances (e.g., any node at equal distance of $i$ and $j$ is considered as closer to $i$ if and only if $i< j$).
Observe that the  space complexity of the algorithm is in $O(m)$, where $m$ is the number of edges of $G$ (i.e., the graph size).

\section{Link with modularity}
\label{sec:mod}

The modularity is a standard metric to assess the quality of a clustering $C$ (any partition of $V$).
Let $\delta_C(i,j) = 1$ if $i,j$ are in the same cluster under clustering  $C$, and $\delta_C(i,j) = 0$ otherwise. The  modularity of clustering $C$ is defined by  \cite{newman2004}:
\be\label{eq:mod}
Q(C) = \frac 1 w \sum_{i,j\in V} (A_{ij}  - \frac{w_iw_j}w)\delta_C(i,j),
\ee
which can be written in terms of probability distributions,
$$
Q(C) =  \sum_{i,j\in V} (p(i,j) - p(i)p(j))\delta_C(i,j).
$$
Thus the modularity is the difference between the probabilities of sampling two nodes of the same cluster under the joint distribution $p(i,j)$ and under the product distribution $p(i)p(j)$.
It can also be expressed from the probability distributions at the cluster level,
$$
Q(C) = \sum_{a\in C} (p(a,a) - p(a)^2).
$$

It is clear from  \eqref{eq:mod} that any clustering $C$ maximizing modularity has some resolution limit, as pointed out in \cite{fortunato2007}, because the second term is normalized by  the total  weight $w$ and thus becomes negligible for too small clusters. To go beyond this resolution limit, it is necessary to introduce a multiplicative factor $\gamma$, called the resolution. The modularity becomes:
\be\label{eq:Qs}
Q_\gamma(C) =  \sum_{i,j\in V} (p(i,j) - \gamma p(i)p(j))\delta_C(i,j),
\ee
or equivalently,
$$
Q_\gamma(C) = \sum_{a\in C} (p(a,a) - \gamma p(a)^2).
$$
This resolution parameter can  be  interpreted  through the Potts model of statistical physics \cite{reichardt2006}, random walks \cite{lambiotte2015}, or statistical inference of a stochastic block model \cite{newman2016}.
For $\gamma = 0$,  the resolution is minimum and 
there is a single cluster, that is
$C= \{\{1,\ldots,n\}\}$; for $\gamma \to +\infty$, the resolution is maximum and 
each node has its own cluster, that is $C= \{\{{1\},\ldots,\{n\}}\}$. 

The Louvain algorithm consists, for any {\it fixed} resolution parameter $\gamma$,  of the following steps:

\begin{enumerate}
\item {\bf Initialization}\\
 $\C \gets \{\{1\} , \ldots , \{n\}\}$
\item {\bf Iteration}\\
 While modularity $Q_\gamma(C)$ increases, update $C$ by moving one node  from one cluster to another.
\item {\bf Aggregation}\\ Merge all nodes belonging to the same cluster, update the  weights and apply step 2 to the resulting aggregate graph while modularity is increased.
\item Return $\C$
\end{enumerate}
The result of step 2 depends on the order in which nodes and clusters are considered; typically, nodes are considered in a cyclic way and the target cluster of each node is  that maximizing the modularity increase. 

Our algorithm can be viewed as a modularity-maximizing scheme with a {\it sliding} resolution. Starting from the maximum resolution where each node has its own cluster, we  look for the first value of the  resolution parameter $\gamma$, say $\gamma_1$, that triggers a single merge between two nodes, resulting in clustering $C_1$. In view of \eqref{eq:Qs}, we have:
$$
\gamma_1 = \max_{i,j\in V} \frac{p(i,j)}{p(i)p(j)}.
$$
These two nodes are merged (corresponding to the aggregation phase of the Louvain algorithm) and we look for the next  value of the  resolution parameter, say $\gamma_2$, that triggers a single merge between two nodes, resulting in clustering $C_2$, and so on. By construction, the 
resolution at time $t$ (that triggers the $t$-th merge) is $\gamma_t = 1 / d_t$ and the corresponding clustering $C_t$ is that of  our algorithm. In particular, the sequence of resolutions $\gamma_1,\ldots,\gamma_{n-1}$ is non-increasing.

To summarize, our algorithm consists of a simple but deep modification of the Louvain algorithm, where 
 the iterative step  (step 2) is replaced by a single merge, at the best current resolution (that resulting in a single merge). 
 In particular, unlike the Louvain algorithm, our algorithm provides a full hierarchy. Moreover, the sequence of resolutions  $\gamma_1,\ldots,\gamma_{n-1}$ can  be used as an input to the Louvain algorithm. 
 Specifically, the resolution $\gamma_t$ provides exactly  $n-t$ clusters in our case, and the Louvain algorithm is expected to provide approximately the same number of clusters at this resolution. 


\section{Experiments}
\label{sec:exp}

We have coded 
our hierarchical clustering algorithm, we refer to as Paris\footnote{Paris $=$ Pairwise  AgglomeRation Induced by Sampling.},  in Python. All  material necessary to reproduce the experiments presented below is available online\footnote{See \href{https://github.com/tbonald/paris}{https://github.com/tbonald/paris}}.

\begin{figure}[h]
\begin{center}
 \subfloat[Level 1]{   \includegraphics[width = 4cm]{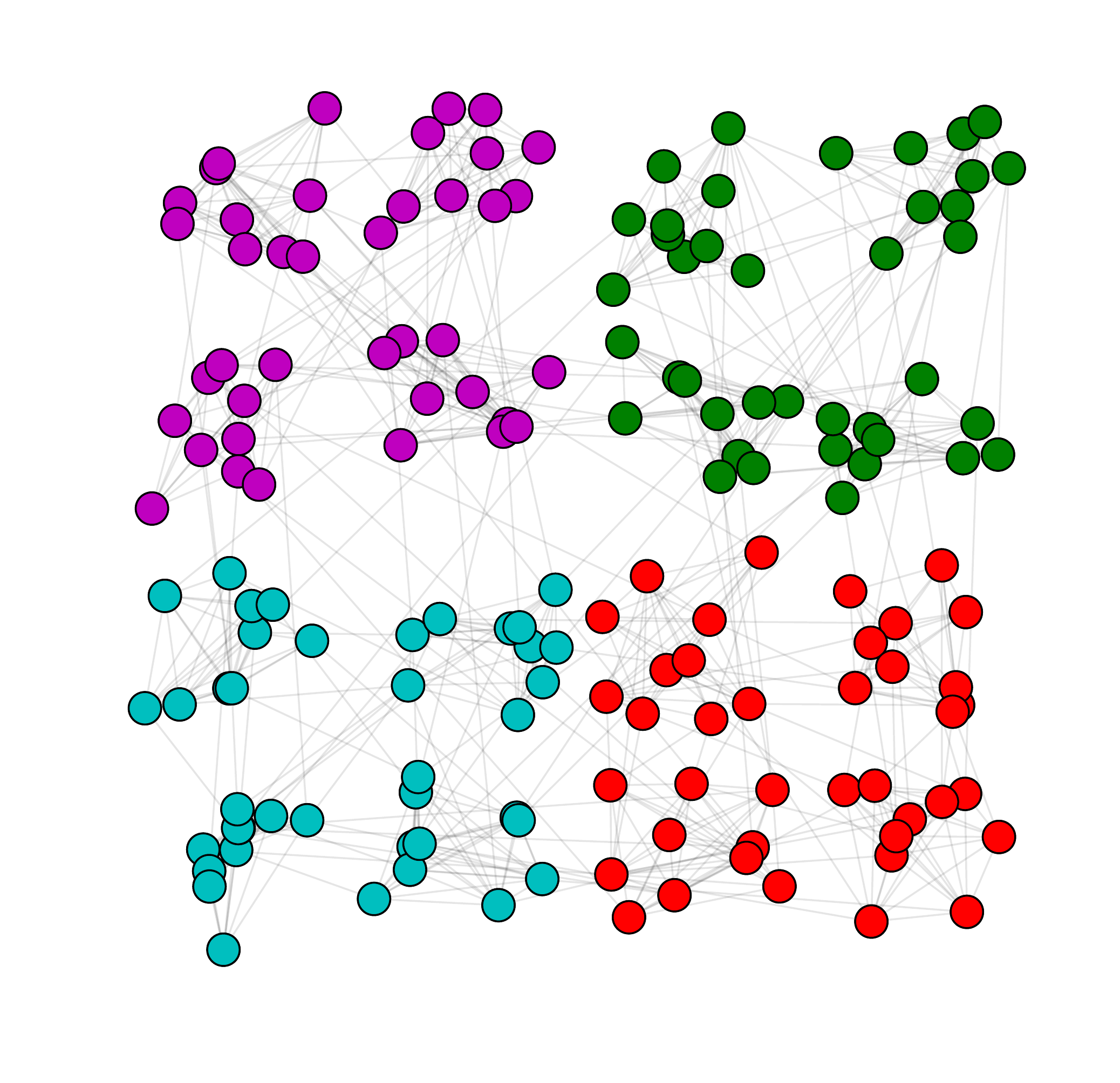}}
\subfloat[Level 2]{    \includegraphics[width = 4cm]{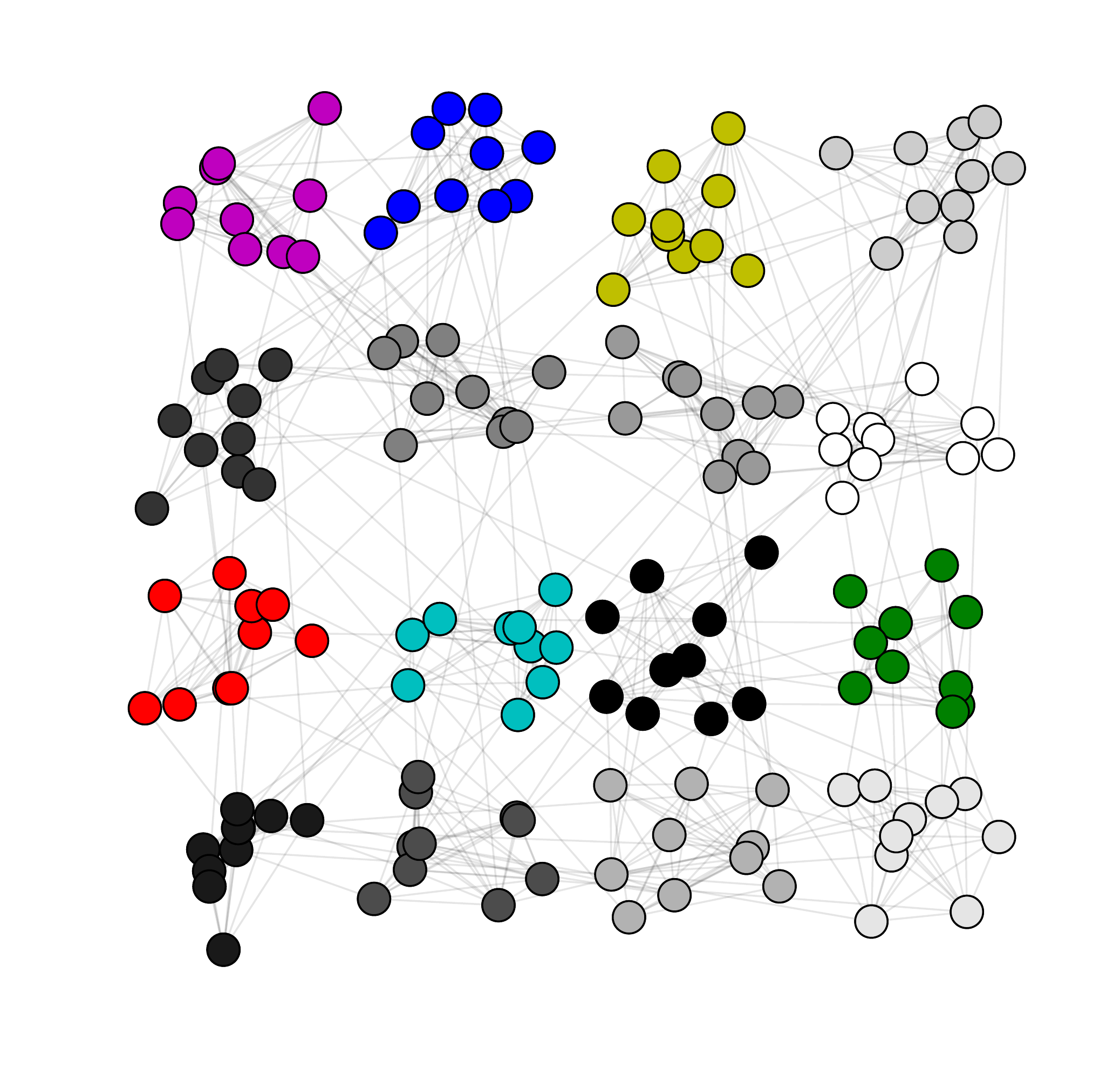}}
    \caption{\label{fig:hsbm-graph}A hierachical stochastic block model with 2 levels of hierarchy.}
    \end{center}
\end{figure}

 \paragraph{Qualitative results.}
We start with  a  hierarchical stochastic block model, as described in \cite{hsbm}. There are $n = 160$ nodes structured in $2$ levels, with $4$ blocks of $40$ nodes at level 1, each  block of 40 nodes being divided into  $4$  blocks of  $10$ nodes at level 2 (see Figure \ref{fig:hsbm-graph}). 


 
\begin{figure}[h]
\begin{center}
    \includegraphics[width = 7cm]{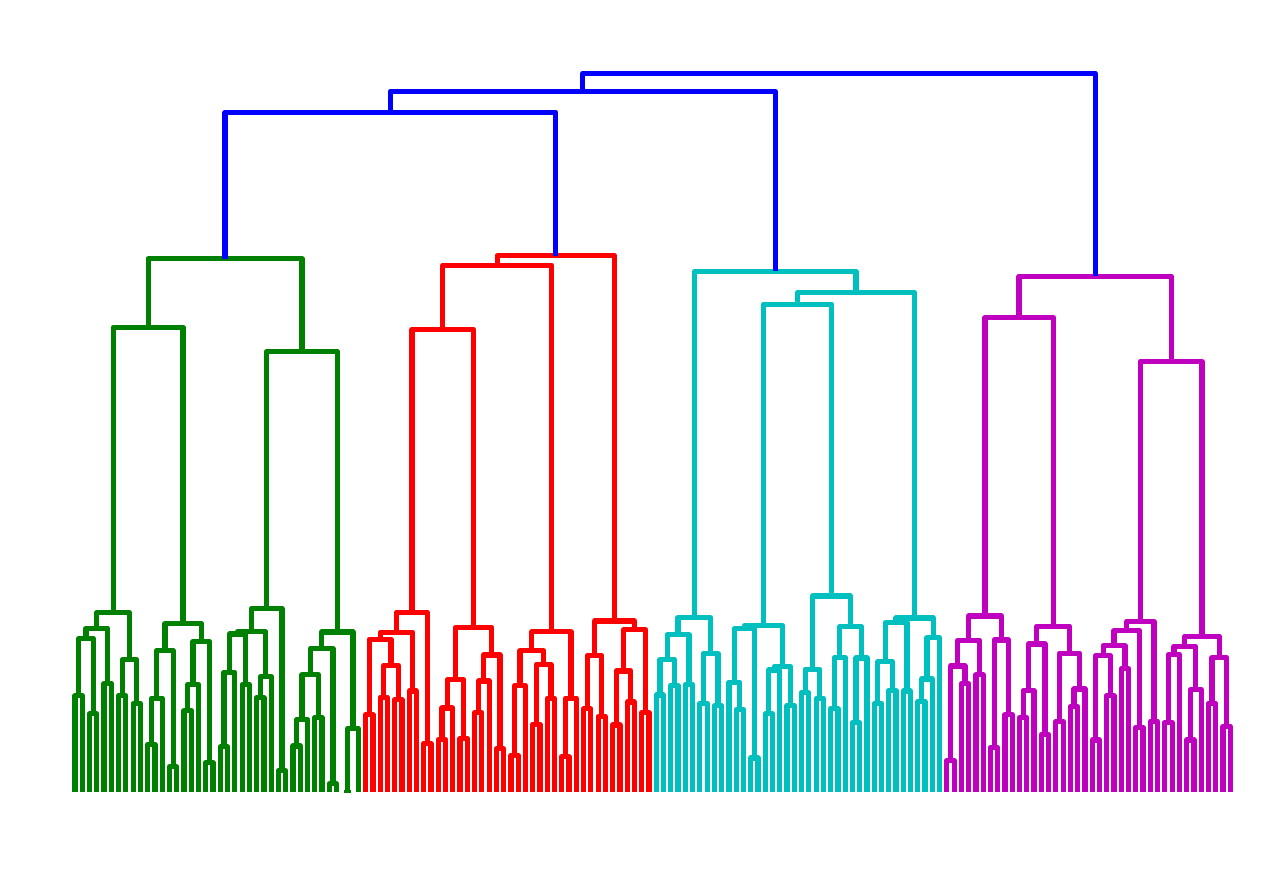}
    \caption{\label{fig:hsbm-dend}Dendrogram associated with the  clustering of Paris on a hierachical stochastic block model of 16 blocks.}
    \end{center}
\end{figure}

The output of Paris is shown in Figure \ref{fig:hsbm-dend} as a dendrogram where the distances (on the $y$-axis) are in log-scale. The two levels of hierarchy clearly appear. 

 

\begin{figure}[h]
\begin{center}
   \includegraphics[width = 8cm]{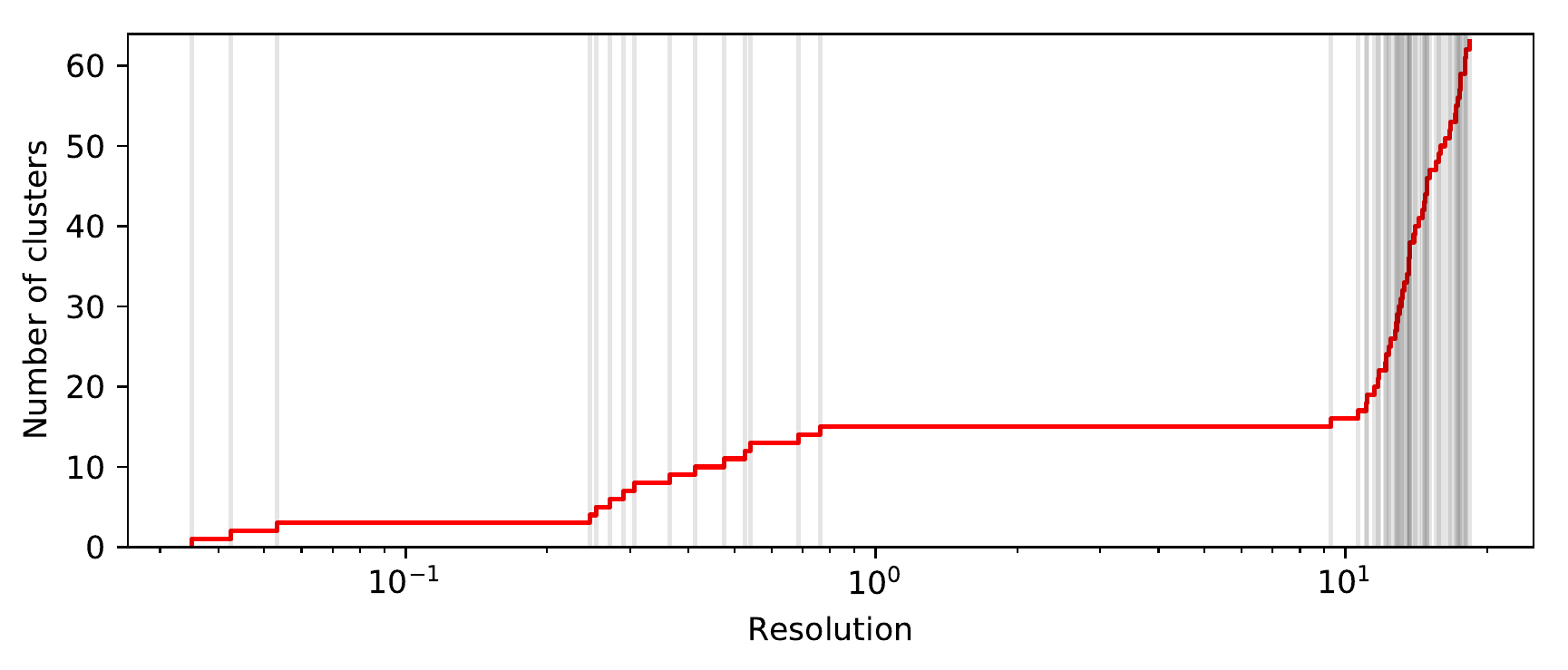}
        \includegraphics[width = 8cm]{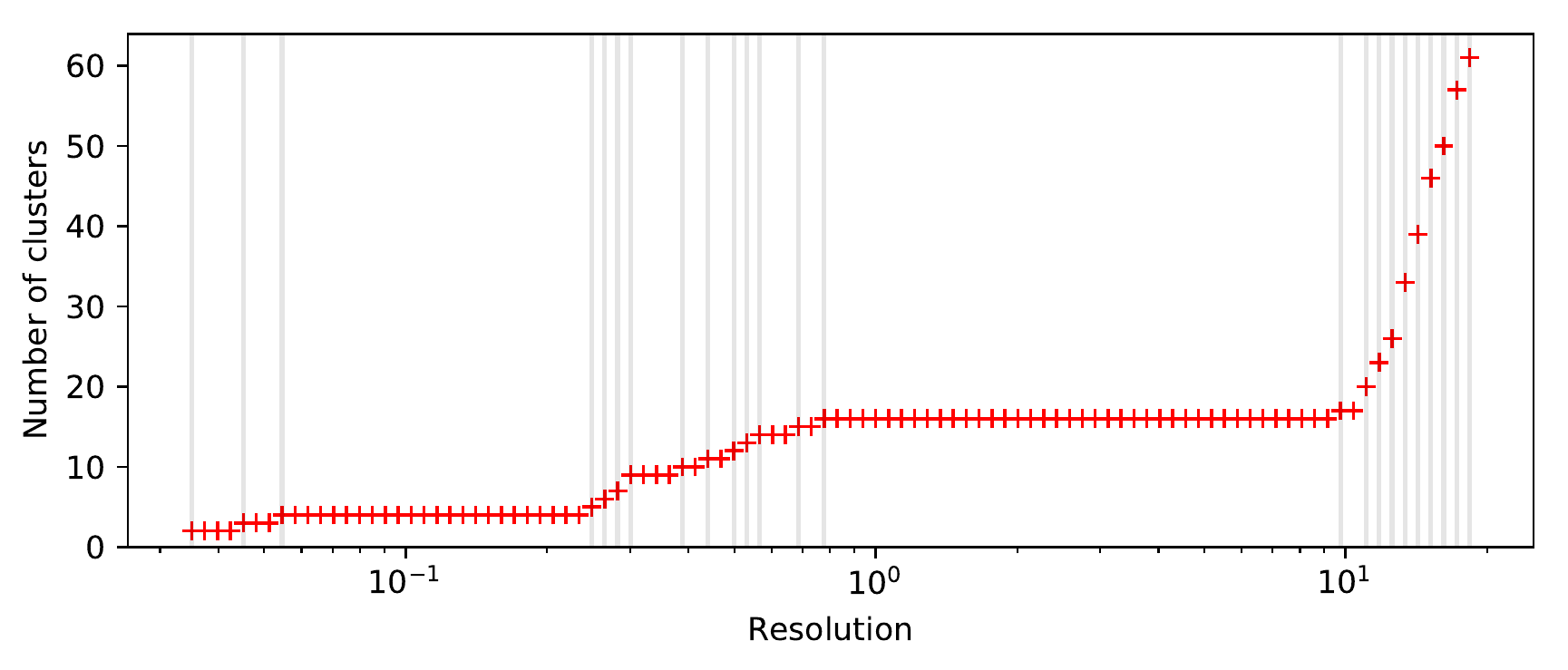}
    \caption{\label{fig:hsbm-res}Number of clusters with respect to the resolution parameter $\gamma$ for Paris (top) and Louvain (bottom) on the hierachical stochastic block model of Figure \ref{fig:hsbm-graph}.}
    \end{center}
\end{figure}

We also show in Figure \ref{fig:hsbm-res} the number of clusters with respect to the resolution parameter $\gamma$ for Paris (top) and Louvain (bottom). The results are very close, and clearly show the hierarchical structure of the model (vertical lines correspond to changes in the number of clusters). The key difference between both algorithms is that, while  Louvain needs to be run for {\it each} resolution parameter $\gamma$ (here 100 values  ranging from $0.01$ to $20$), 
Paris is run only once, the relevant resolutions being direct outputs of  the algorithm, embedded in the dendrogram (see Section \ref{sec:mod}). 

We now consider four  real datasets, whose characteristics are summarized in Table \ref{tab:dataset}.

 \begin{table}[h]
\begin{center}
\begin{tabular}{l|c|c|c}
Graph & \# nodes  & \# edges & Avg. degree\\
\hline
OpenStreet      &    5,993  &  6,957 &  2.3 \\
OpenFlights      &    3,097 &  18,193  & 12 \\
Wikipedia Schools &     4,589 & 106,644 & 46 \\
Wikipedia Humans &     702,782 & 3,247,884 & 9.2
\end{tabular}
\end{center}
\caption{Summary of the datasets.}
\label{tab:dataset}
\end{table}

\begin{figure}[ht!]
\begin{center}
   \includegraphics[width = 9cm]{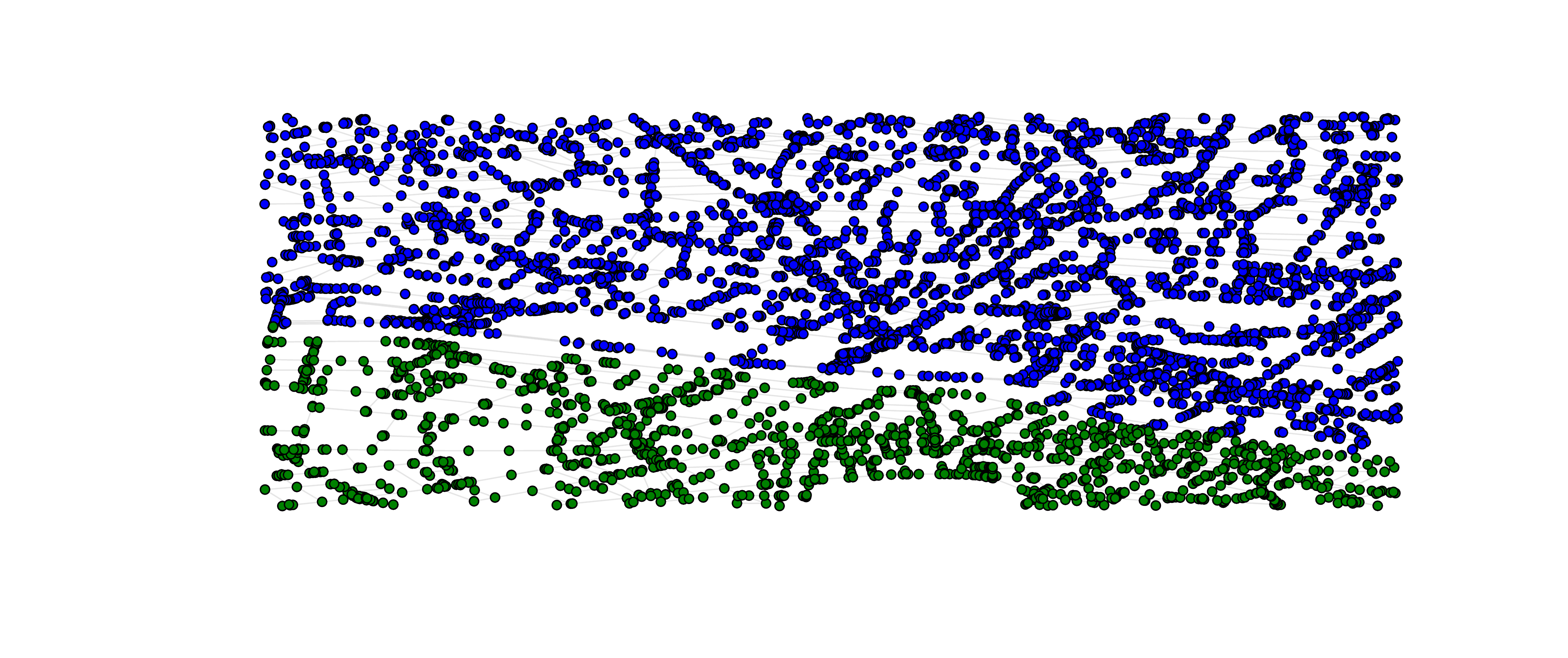}
    \includegraphics[width = 9cm]{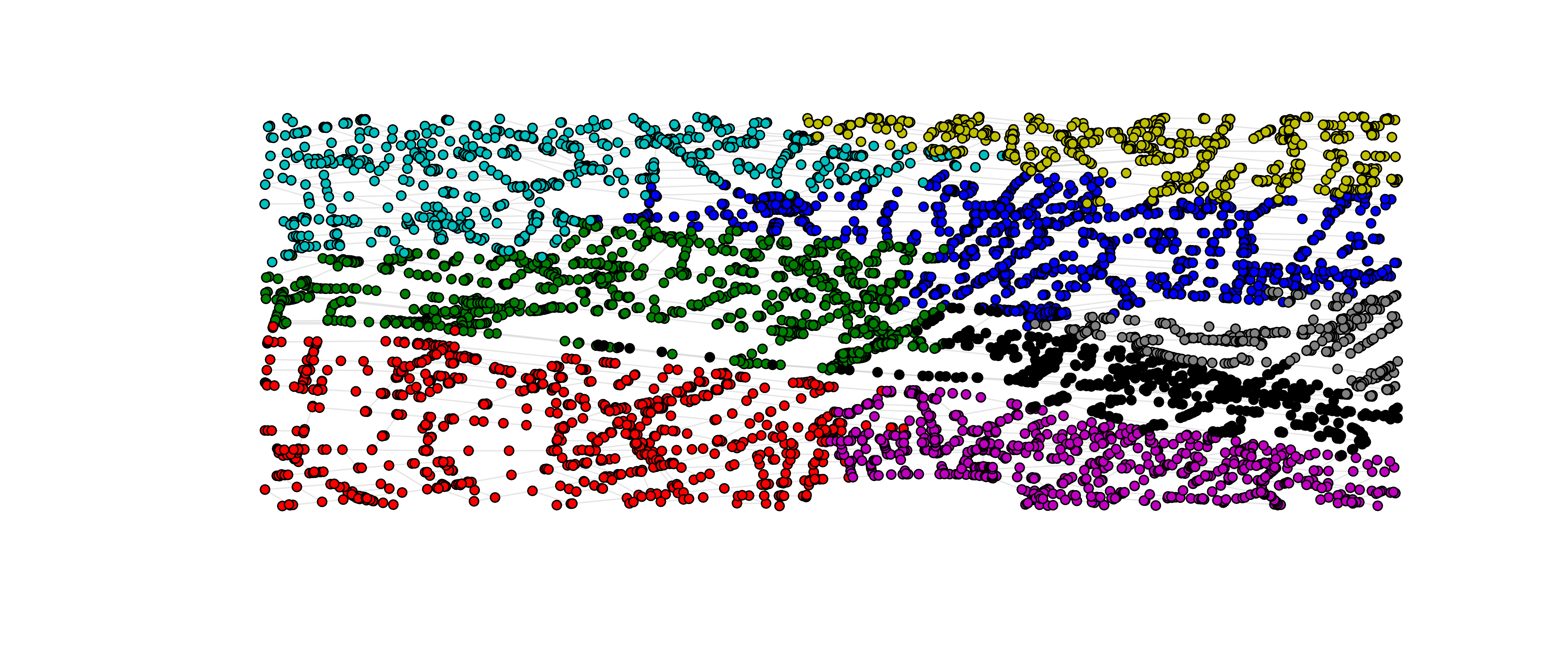}
    \caption{\label{fig:openstreet}Clusterings of the OpenStreet graph by Paris.}
    \end{center}
\end{figure}

\vspace{-.5cm}
The first dataset, extracted from OpenStreetMap\footnote{https://openstreetmap.org},  is the graph formed by the streets of the center of Paris. To illustrate the quality of the hierarchical clustering returned by our algorithm, we have extracted the two ``best" clusterings, in terms of ratio between successive distance merges in the corresponding dendrogram; the results  are shown in Figure \ref{fig:openstreet}. The best clustering gives two clusters, Rive Droite (with Ile de la Cit\'e) and Rive Gauche, the two banks separated by the river Seine; the second best clustering divides these two clusters into sub-clusters.

\clearpage

\begin{figure}[ht!]
\begin{center}
 \includegraphics[width = 16cm]{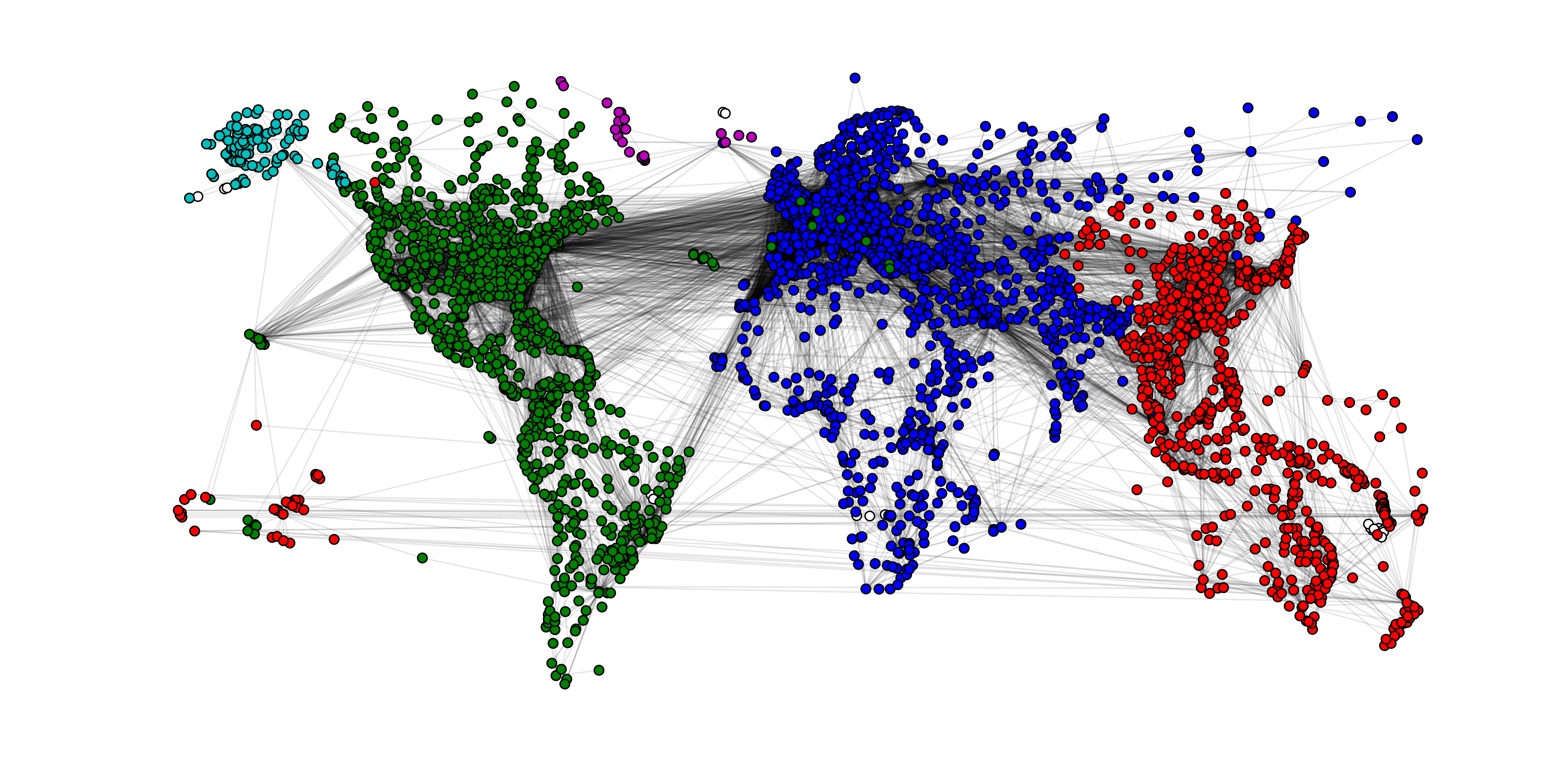}
    \includegraphics[width = 16cm]{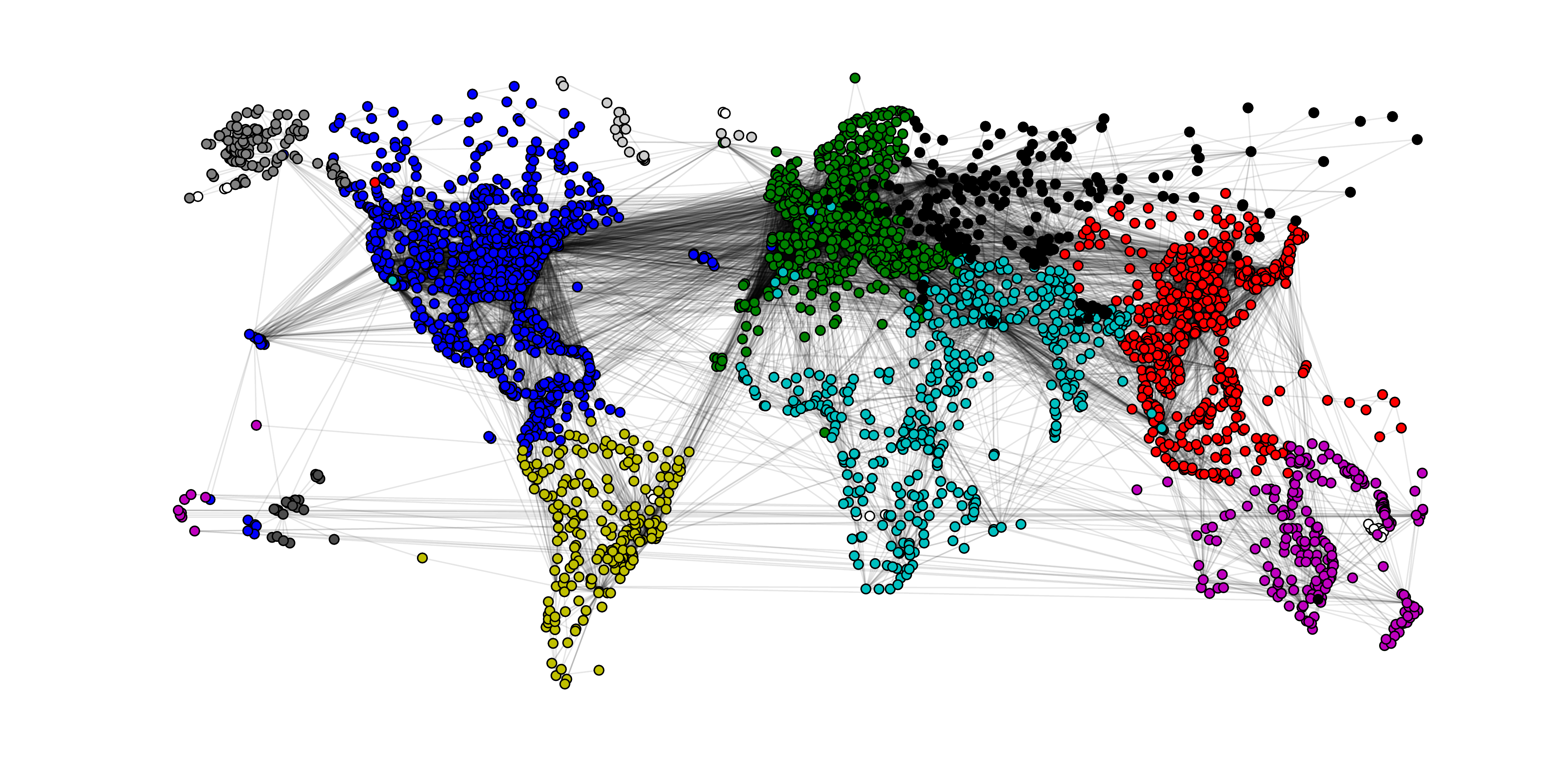}
    \caption{\label{fig:openflight}Clusterings of the OpenFlights graph by Paris.}
    \end{center}
\end{figure}

The second dataset, extracted from OpenFlights\footnote{https://openflights.org}, is the  graph of airports with the weight between two airports equal to the number of daily flights between them. We run Paris and extract the best clusterings from the largest component of the graph, as for the OpenStreet graph. The first two best clusterings  isolate the Island/Groenland area and the Alaska from the rest of the world, the corresponding airports forming dense clusters, lightly connected with the other airports. The following two best clusterings are shown in Figure \ref{fig:openflight}, with respectively 5 and 10 clusters corresponding to meaningful continental  regions of the world.

\clearpage
 
 The third dataset is the graph formed by links between pages of 
Wikipedia for Schools\footnote{https://schools-wikipedia.org}, see \cite{west2009,west2012}. 
The graph is considered as undirected.
Table \ref{tab:wiki} (top) shows  the 10 largest clusters  of $C_{n-100}$,  the 100 last  (and thus largest) clusters  found by Paris. Only   pages of highest degrees are shown for   each cluster.

\vspace{1cm}

 \begin{table}[h]
\begin{center}
The 10 largest clusters of Wikipedia for Schools among 100 clusters found by Paris
\begin{tabular}{|l|l|}
\hline
Size & Main pages\\
\hline
288&
Scientific classification, Animal, Chordate, Binomial nomenclature, Bird \\

231&
Iron, Oxygen, Electron, Hydrogen, Phase (matter) \\

196&
England, Wales, Elizabeth II of the United Kingdom, Winston Churchill, William Shakespeare \\

164&
Physics, Mathematics, Science, Albert Einstein, Electricity\\

148&
Portugal, Ethiopia, Mozambique, Madagascar, Sudan \\

139&
Washington, D.C., President of the United States, American Civil War, Puerto Rico, Bill Clinton\\

129&
Earth, Sun, Astronomy, Star, Gravitation \\

127&
Plant, Fruit, Sugar, Tea, Flower \\

104&
Internet, Computer, Mass media, Latin alphabet, Advertising\\

99&
Jamaica, The Beatles, Hip hop music, Jazz, Piano \\
\hline
\end{tabular}
\end{center}
\begin{center}
The 10 largest subclusters of the first cluster above
\begin{tabular}{|l|l|}
\hline
Size & Main pages\\
\hline
71&
Dinosaur, Fossil, Reptile, Cretaceous, Jurassic \\

51&
Binomial nomenclature, Bird, Carolus Linnaeus, Insect, Bird migration \\

24&
Mammal, Lion, Cheetah, Giraffe, Nairobi \\

22&
Animal, Ant, Arthropod, Spider, Bee \\

18&
Dog, Bat, Vampire, George Byron, 6th Baron Byron, Bear \\

16&
Eagle, Glacier National Park (US), Golden Eagle, Bald Eagle, Bird of prey\\

16&
Chordate, Parrot, Gull, Surtsey, Herring Gull \\

15&
Feather, Extinct birds, Mount Rushmore, Cormorant, Woodpecker \\

13&
Miocene, Eocene, Bryce Canyon National Park, Beaver, Radhanite \\

12&
Crow, Dove, Pigeon, Rock Pigeon, Paleocene\\
\hline
\end{tabular}
\end{center}
\caption{Hierarchical clustering of the graph  Wikipedia for Schools by Paris.}
\label{tab:wiki}
\end{table}


Observe that the ability of selecting the clustering associated with some target number of clusters  is one of the key advantage of Paris over Louvain.  Moreover, Paris gives a full hiearchy of the pages, meaning that each of these clusters is divided into sub-clusters in the output of the algorithm. 
 Table \ref{tab:wiki} (bottom) gives for instance, among the 500 clusters found by Paris (that is,  in $C_{n-500}$),   the 10 largest clusters that are subclusters of the first cluster of Table \ref{tab:wiki} (top), related to animals / taxonomy. The subclusters tend to give meaningful groups of animals, revealing the multi-scale structure of the dataset.

The fourth dataset is the subgraph of Wikipedia restricted to pages related to humans. We have done  the same experiment as for Wikipedia for Schools. The results are shown in Table \ref{tab:wikihumans}.
Again, we observe that clusters form relevant groups of people, with cluster \#1 corresponding to political figures for instance, this cluster consisting of meaningful subgroups as shown in the bottom of Table \ref{tab:wikihumans}. All this information is embedded in the dendrogram returned by Paris.
\clearpage

 \begin{table}[h]
\begin{center}
The 10 largest clusters of Wikipedia Humans among 100 clusters found by Paris
\begin{tabular}{|l|l|}
\hline
Size & Main pages\\
\hline
41363&
George W. Bush, Barack Obama, Bill Clinton, Ronald Reagan, Richard Nixon \\

34291&
Alex Ferguson, David Beckham,  Pel\'e, Diego Maradona, Jos\'e Mourinho \\

25225&
Abraham Lincoln, George Washington, Ulysses  S. Grant, Thomas Jefferson, Edgar Allan Poe \\

23488&
Madonna, Woody Allen,  Martin Scorsese, Tennessee Williams, Stephen Sondheim \\

23044&
Wolfgang Amadeus Mozart, Johann Sebastian Bach,  Ludwig van Beethoven, Richard Wagner \\

22236&
Elvis Presley, Bob Dylan,  Elton John, David Bowie, Paul McCartney\\

20429&
Queen Victoria, George III of the UK, Edward VII,  Charles Dickens, Charles, Prince of Wales \\

19105&
Sting, Jawaharlal Nehru,  Rabindranath Tagore, Indira Gandhi, Gautama Buddha\\
18348&
Edward I of England, Edward III of England,   Henry II of England, Charlemagne \\

14668&
Jack Kemp, Brett Favre,   Peyton Manning, Walter Camp, Tom Brady \\
\hline
\end{tabular}
\end{center}
\begin{center}
The 10 largest subclusters of the first cluster above.
\begin{tabular}{|l|l|}
\hline
Size & Main pages\\
\hline
2722&
Barack Obama, John McCain, Dick Cheney, Newt Gingrich, Nancy Pelosi \\

2443&
Arnold Schwarzenegger, Jerry Brown, Ralph Nader, Dolph Lundgren, Earl Warren\\

2058&
Osama bin Laden, Hamid Karzai, Alberto Gonzales, Janet Reno,  Khalid Sheikh Mohammed \\

1917&
Dwight D. Eisenhower, Harry S. Truman, Douglas MacArthur, George S. Patton\\

1742&
George W. Bush, Condoleezza Rice, Colin Powell, Donald Rumsfeld, Karl Rove \\

1700&
Bill Clinton, Thurgood Marshall, Mike Huckabee, John Roberts, William Rehnquist\\

1559&
Ed Rendell, Arlen Specter, Rick Santorum, Tom Ridge, Mark B. Cohen\\

1545&
Theodore Roosevelt, Herbert Hoover, William Howard Taft, Calvin Coolidge\\

1523&
Ronald Reagan, Richard Nixon, Jimmy Carter, Gerald Ford, J. Edgar Hoover\\

1508&
Rudy Giuliani, Michael Bloomberg,  Nelson Rockefeller, George Pataki, Eliot Spitzer\\
\hline
\end{tabular}
\end{center}

\caption{Hierarchical clustering of the graph  Wikipedia Humans by Paris.}
\label{tab:wikihumans}
\end{table}

\paragraph{Quantitative results.} To assess the quality of the hierarchical clustering, we use the cost function proposed in \cite{dasgupta2016cost} and given by:
\be\label{eq:cost}
\sum_{a,b} p(a,b) (|a| + |b|),
\ee
where the sum is over all left and right clusters $a,b$ attached to any internal node of the  tree representing the hierarchy. This is the expected size of the smallest subtree  containing two random nodes $i,j$,  sampled from the joint distribution $p(i,j)$ introduced in Section \ref{sec:dist}. If the tree indeed reflects the underlying hierarchical structure of the graph, we expect most edges to link nodes that are close in the tree, i.e., whose  common ancestor is relatively far from the root. The size of the corresponding subtree (whose root is this common ancestor) is expected to be small, meaning that  \eqref{eq:cost} is a relevant cost function.  
Moreover, it was proved in \cite{cohen2018hierarchical} that if the graph is perfectly hierarchical, the underlying tree is optimal with respect to this cost function. 

The results are presented in Table \ref{tab:perf} for  the  graphs considered so far and  the graphs of Table \ref{tab:SNAP}, selected from the SNAP datasets  \cite{snapnets}.
The cost function is normalized by the   number of nodes $n$ so as to get a value between 0 and 1. 
We compare the performance of Paris to that of a  spectral algorithm where the nodes are embedded in a space of dimension 20  
by using the  20 leading eigenvectors of the Laplacian matrix $L = D - A$ ($D$ is the diagonal matrix of node weights) and applying 
 the Ward 
method in the embedding space.
The spectral decomposition of the Laplacian is based on standard functions on sparse matrices available in the  Python package {\tt scipy}.

Observe that we do not include Louvain in these experiments as this algorithm does not provide a full hierarchy of the graph, so that the cost function \eqref{eq:cost} is not applicable.

 \begin{table}[h]
\begin{center}
\begin{tabular}{l|c|c|c}
Graph & \# nodes  & \# edges & Avg. degree\\
   \hline
   Facebook & 4,039 & 88,234 & 44\\
   Amazon & 334,863 & 925,872 & 5.5\\
   DBLP & 317,080 & 1,049,866 & 6.6\\
   Twitter & 81,306 & 1,342,310 & 33\\
   Youtube & 1,134,890 & 2,987,624 & 5.2\\
   Google & 855,802 & 4,291,352 & 10
\end{tabular}

\end{center}
\caption{Summary of the considered graphs from SNAP.}
\label{tab:SNAP}
\end{table}

The results  are shown when the algorithm runs within some time limit (less than 1 hour on a computer equipped with a 2.8GHz Intel Core i7 CPU with 16GB of RAM). The best performance is displayed in bold characters.
Observe that both algorithms have similar performance on those graphs where the results are available.  However, Paris is much faster than the spectral algorithm, as shown by Table \ref{tab:run} (for each algorithm, the initial load or copy of the graph is not included in the running time; running times exceeding 1 hour are not shown). Paris is even faster than Louvain in most cases, while providing a much richer information on the graph.

 \begin{table}[h]
\begin{center}
\begin{tabular}{l|c|c}
Graph  & Spectral     &   Paris \\
\hline
OpenStreet    & 0.0103   & {\bf 0.0102}   \\
OpenFlights    &  {\bf 0.125}  & {0.130} \\
 Facebook  & 0.0479  & \textbf{0.0469}\\
Wikipedia Schools&   0.452    &  {\bf 0.402}\\
   Amazon  & $-$  & \textbf{0.0297}\\
   DBLP  & $-$  & \textbf{0.110}\\
   Twitter & $-$   & {\bf 0.0908}\\
   Youtube &  $-$  &  {\bf 0.185}\\
Wikipedia Humans &  $-$  &  {\bf 131}  \\
 Google & $-$ & {\bf 0.0121}
\end{tabular}
\end{center}
\caption{Performance comparison of a spectral algorithm and Paris  in terms of  normalized Dasgupta's cost.}
\label{tab:perf}
\end{table}


 \begin{table}[h]
\centering
\begin{tabular}{l|c|c|c}
Graph &    Spectral  & Louvain & Paris \\
   \hline
   OpenStreet  & 0.86s & 0.19s & \textbf{0.17s}\\
   OpenFlight  & 0.51s& \textbf{0.31s}& 0.33s \\
   Facebook & 5.9s& 1s & \textbf{0.71s} \\
   Wikipedia Schools  & 16s  & 2.2s & \textbf{1.5s}\\   
   Amazon &  $-$ & 45s & \textbf{43s} \\
   DBLP &  $-$ & 52s & \textbf{31s} \\
   Twitter &  $-$ & 35s & \textbf{21s} \\
   Youtube &  $-$ & \textbf{8 min} & 16 min 30s \\
   Wikipedia Humans  & $-$ & 2 min 30s & \textbf{2 min 10s}\\
   Google &  $-$ & 3 min 50s & \textbf{1 min 50s} 
\end{tabular}
\caption{\label{tab:run}
Comparison of running times.}
\end{table}

\section{Conclusion}
\label{sec:conc}

We have proposed an agglomerative  clustering algorithm for graphs based on a reducible distance between clusters. 
The algorithm  captures the multi-scale structure of real graphs. It is parameter-free, fast and memory-efficient.
In the future, we plan to work on the automatic extraction of   clusterings from the dendrogram, at the most relevant resolutions.


\end{document}